\documentclass[aps,pra,reprint,amsmath,amssymb,graphicx,longbibliography]{revtex4-2}

\usepackage{bm}
\usepackage{amsmath}
\usepackage{amsfonts}
\usepackage[normalem]{ulem}
\usepackage[colorlinks=true,allcolors=blue,bookmarks=false,pdfusetitle]{hyperref}
\usepackage{url}
\usepackage{xcolor} 
\usepackage{graphicx}
\usepackage{breqn}
\usepackage{braket}
\usepackage{cases}
\usepackage[utf8]{inputenc}
\usepackage[T1]{fontenc}

\usepackage[overload]{empheq}
\usepackage{cleveref}

\usepackage{tikz}

\makeatletter
\let\cat@comma@active\@empty
\makeatother

\begin{document}

\title{
Universal principles for sudden-quench quantum Otto engines }

\author{R. S. Watson}
\affiliation{School of Mathematics and Physics, University of Queensland, Brisbane,  Queensland 4072, Australia}
\author{K. V. Kheruntsyan}
\affiliation{School of Mathematics and Physics, University of Queensland, Brisbane, Queensland 4072, Australia}

\date{\today{}}

\begin{abstract}
\noindent 
 We apply a simple sudden quench approximation for the unitary work strokes of a quantum Otto engine in order to provide a general analysis of its performance, applicable to arbitrary quantum models with two-body interactions.
This work extends recent results for an interaction-driven Otto cycle to generic many-body interacting quantum models, providing universal bounds on their operation efficiency. From this, we demonstrate that the net work of such an engine cycle is determined entirely by interparticle correlations. Applications are demonstrated for a handful of paradigmatic many-body quantum models, including a novel engine---with a spin-1/2 Fermi gas with contact two-body interactions as its working medium---in which we leverage control over spin polarization to greatly enhance its performance compared to the unpolarized case.
We then extend the analysis of interaction-driven quantum Otto engine cycles to systems where control is exerted over the strength of arbitrary quantum operators that might be present in the system Hamiltonian (such as one-body, or three-body, etc.), finding that the general principles derived for the sudden quench with two-body interactions apply universally. As an example, this is demonstrated for a conventional volumetric Otto cycle, where beneficial net work is generated by leveraging the control over the frequency of an external trap, which is a one-body operator.
However, we emphasize that the results derived here apply universally to all Otto engine cycles operating under a sudden quench protocol.
\end{abstract}

\maketitle

\section{Introduction}

The universal utility of thermodynamics is due in large part to its ability to make general predictions without specific knowledge of the underlying model, which may be described classically or quantum mechanically \cite{CallenHerbertB1985Taai,SchroederD_ThermalPhysics,bender2000,gardas2015carnot,xu2018carnot,Bera2022carnot}. Indeed, the first and second laws of thermodynamics were formulated over 50 years prior to the discovery of the atom \cite{pullman2001atom}. Further, in his original analysis of the heat engine, Sadi Carnot provided the correct universal upper bound on its efficiency with no recourse to an underlying microscopic dynamics \cite{carnot1890reflections}.

The modern theory of quantum thermodynamics is motivated by the study of thermodynamic phenomena in a quantum context \cite{kosloff2013quantum,kosfloff2014quantum,alicki2018introduction,anders2017focus,vinjanampathy2016quantum,deffner2019quantum,Myers2022quantum,millen2016perspective,potts2024quantum,Bhattacharjee2021,Gluza2021}. In particular, the past decade has seen a surge in interest towards realizing thermodynamic devices and protocols within quantum mechanical systems, an interest that has been motivated in large part by the rapid advancement in experimental methods and techniques, which have allowed for unprecedented levels of control over quantum platforms \cite{Rossnagel2016single,Ca-ion-spin-engine,Maslennikov2019,VanHorne2020,Nitrogen-vacancy-heat-engine,bouton2021quantum}, and the prospect of gaining a `quantum advantage' over the classical counterparts of such devices and protocols
\cite{funo2013thermo,alicki2013entanglement,hilt2009system,llobet2015extractable,Huber_2015,narasimhachar2015low,Korzekwa2016extraction,jaramillo2016quantum,halpern2019quantum}.

Precise experimental control has recently allowed for the experimental realization of uniquely quantum \emph{many-body} engine cycles operating in the quasi-static limit \cite{koch2022making,simmons2023thermodynamic}. However, one may instead consider nonequilibrium quantum engine cycles, which naturally operate at finite power, in contrast to quasi-static operation which operates with maximum efficiency, but vanishingly small power. In such cases, it is natural to ask whether it is possible to make universal statements on the performance of such nonequilibrium engine cycles, as one is able to for quasi-static operation \cite{Kosloff2017quantum,Bera2022carnot,vinjanampathy2016quantum}.

\begin{figure}[!b]
\centering
\includegraphics[width=0.47\textwidth]{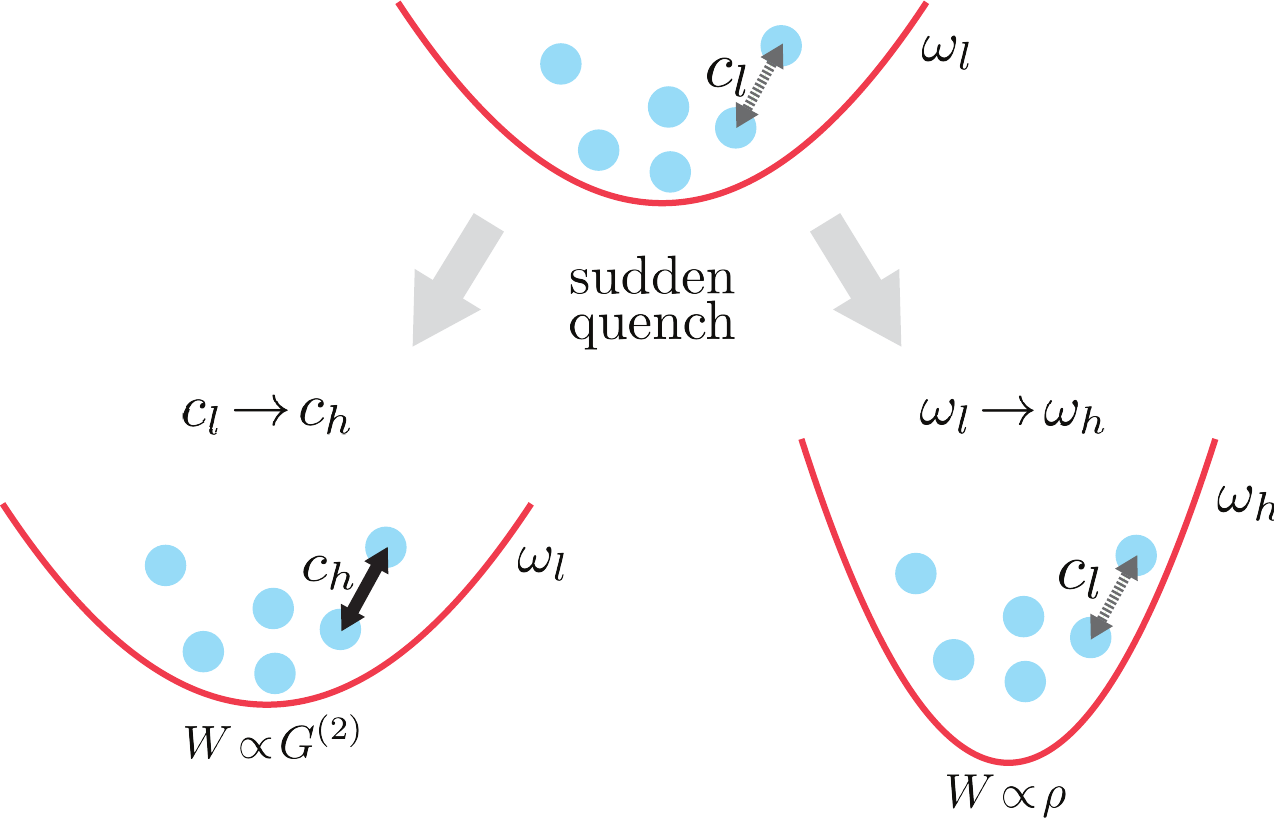}
\caption{Cartoon illustration of the sudden-quench work strokes in the quantum Otto engine cycle investigated in this work. The net work of such a cycle depends on the parameter chosen to quench, illustrated here for an interacting quantum gas in an external harmonic trap, for two important cases: \emph{interaction} driving (left), where control is over interaction strength, $c$, and the net work is determined by interparticle correlations, $G^{(2)}$; and \emph{volumetric} driving (right), where control is over the external trap frequency, $\omega$, and the net work is determined by the atomic density distribution, $\rho$.}
\label{fig:interaction_Engine_Diagram}
\end{figure}

In this work, we utilize a sudden quench approximation, which greatly simplifies the resulting dynamics, to provide a general analysis of the performance of nonequilibrium quantum Otto engines.
In particular, in Section \ref{sec:Otto} we extend the recent analysis on many-body quantum Otto engine cycles operating under a sudden quench of interaction strength, done for the uniform one-dimensional (1D) Bose gas in Ref.~\cite{watson2025quantum}, to arbitrary quantum many-body interacting models. From this, we demonstrate that the net work of such an interaction-driven Otto engine cycle is universally determined by interparticle two-body correlations.

We formulate general bounds on efficiency of the interaction-driven Otto engine in Section \ref{sec:bounds}. These bounds rely only on the ratio of the interaction strength parameters which are suddenly quenched, regardless of the particular realization in a physical model. We further demonstrate that engine performance in the limiting case of an infinitesimal interaction quench coincides precisely with that of the same cycle performed quasi-statically.

Application of this interaction-driven quantum Otto engine cycle is examined in paradigmatic quantum models in Section \ref{sec:applications_interaction}. There, we extend our previous analysis of the sudden quench quantum Otto engine cycle in the Lieb-Liniger model of a uniform 1D Bose gas, presented in Ref.~\cite{watson2025quantum}, to include the effects of diffusive system-reservoir contact \cite{SchroederD_ThermalPhysics}. This provides an illustrative example in regards to the utility of our newly derived upper bound on efficiency. 
We then investigate an interaction-driven quantum Otto cycle within a harmonically trapped ultracold Bose gas in $1D$, $2D$, and $3D$ where performance is shown to be maximal in the 1D system. 
We additionally introduce a novel Otto engine cycle in an interacting 1D Fermi gas described by the Yang-Gaudin model, where control over spin polarization may be leveraged to greatly enhance performance.

Finally, in Section \ref{sec:generalise_otto} we show how the formalism developed for the interaction-driven cycle in Section \ref{sec:Otto} may be simply generalized to Otto engine cycles where external control is over the strength of any arbitrary quantum operator present in the system Hamiltonian. We apply this general analysis to the operation of a conventional volumetric Otto cycle, with external control over the harmonic trapping frequency of an ultracold Bose gas, demonstrating improved performance with increasing dimension, in contrast to what was seen in the interaction-driven cycle.

\section{Interaction-driven sudden quench Otto engine}\label{sec:Otto}

In the following subsections, we define an interaction-driven quantum many-body Otto cycle, extending the work done previously in Ref.~\cite{watson2025quantum} to arbitrary interacting quantum many-body models.
From this we demonstrate that the net work is determined by the difference of atom-atom correlations measured in the two equilibrium states of our Otto cycle. We then provide a general upper-bound on engine efficiency, and relate the operation of an infinitesimal sudden quench to adiabatic engine operation.

\subsection{Interaction strength quench}
\label{sec:quench}

We begin by introducing the general form of the Hamiltonian for our interacting quantum many-body system that will take the role of the working fluid in the Otto engine cycle. In particular, we separate the total system Hamiltonian, $\hat{H}$, into a sum of two operators,
\begin{equation}\label{eq:hamiltonian}
    \hat{H} = \hat{H}_0 + c \, \hat{G}_2,
\end{equation}
where the operator $\hat{H}_0$ captures the set of terms that are kept constant during the unitary work strokes of the engine (see below). This may correspond to kinetic energy or, more generally, any other terms that are not externally manipulated during the work strokes, such as a static external trapping potential.
The operator $\hat{G}_2$, on the other hand, describes the particular form of interparticle interactions  and is the key quantity of interest throughout this and the following section, whereas the constant $c$ characterizes the strength of interactions.

For many paradigmatic many-body models, the interparticle interaction operator  $\hat{G}_{2}$ can be expressed as an integral or sum over creation, annihilation, or spin operators, such as:
\begin{subequations}
	\label{eq:def}
	\begin{align}[left ={ \hat{G}_{2} \!=\! \empheqlbrace}]
		& \!\iint \!  d\mathbf{r} d\mathbf{r}' \hat{\Psi}^{\dagger}(\mathbf{r}) \hat{\Psi}^{\dagger}(\mathbf{r}') f(\mathbf{r},\mathbf{r}') \hat{\Psi}(\mathbf{r}') \hat{\Psi}(\mathbf{r}),\label{eq:bosons} \\
		& \!\iint\!  d\mathbf{r} d\mathbf{r}' \hat{\Psi}_{\uparrow}^{\dagger}(\mathbf{r}) \hat{\Psi}_{\downarrow}^{\dagger}(\mathbf{r}') f(\mathbf{r}\!-\!\mathbf{r}') \hat{\Psi}^{\,}_{\downarrow}(\mathbf{r}') \hat{\Psi}^{\,}_{\uparrow}(\mathbf{r}),\label{eq:fermions}\\
		& \sum_{j}\hat{S}_j^z\hat{S}_{j+1}^z,\label{eq:Heisenberg} \\
        & \sum\limits_{j} \hat{c}^{\dagger}_{j,\uparrow} \hat{c}^{\dagger}_{j,\downarrow} \hat{c}^{\,}_{j,\downarrow}\hat{c}^{\,}_{j,\uparrow} ,\label{eq:FermiHubbard}\\
		& \sum\limits_{j}  \hat{b}^{\dagger}_j \hat{b}^{\dagger}_j \hat{b}_j^{\,} \hat{b}_j^{\,}, \label{eq:BoseHubbard} 
	\end{align}
\end{subequations}
Equation~\eqref{eq:bosons} describes the interactions in a gas of identical bosons, annihilated (created) at position $\mathbf{r}$  by the field operator $\hat{\Psi}^{(\dagger)}(\mathbf{r})$, with $f(\mathbf{r},\mathbf{r}')$ describing the spatial dependence of the interaction. Similarly, Eq.~\eqref{eq:fermions} describes a system of fermions with spin components up ($\uparrow$) and down ($\downarrow$), annihilated and created by their respective field operators. Equation \eqref{eq:Heisenberg} describes an Ising interaction between spins on a lattice,  $\hat{S}^z_j$.
Lastly, equations \eqref{eq:FermiHubbard} and \eqref{eq:BoseHubbard} are the Fermi-Hubbard and Bose-Hubbard interaction operators, respectively, describing on-site interparticle interactions.

As an illustrative example, the Lieb-Liniger model of a one-dimensional (1D) Bose gas assumes contact interactions, $f(x,x')=\delta(x-x')$ (cf.~Eq.~\eqref{eq:bosons}). 
For a uniform system of length $L$, the interaction energy operator, $\hat{H}_{\text{int}}=c\,\hat{G}_2$, is proportional to
\begin{align}
\label{eq:g2_LL}
   \hat{G}_2&= \int_0^L dx \hat{\Psi}^{\dagger}(x) \hat{\Psi}^{\dagger}(x) \hat{\Psi}(x) \hat{\Psi}(x)\\
    &= L\,\hat{\Psi}^{\dagger}(x) \hat{\Psi}^{\dagger}(x) \hat{\Psi}(x) \hat{\Psi}(x),
\end{align}
such that its expectation value is proportional to the local (same point), unnormalized, two-particle correlation function,
\begin{align}
\label{eq:corr-fn}
   \langle \hat{G}_2\rangle=  L\,\langle \hat{\Psi}^{\dagger}(x) \hat{\Psi}^{\dagger}(x) \hat{\Psi}(x) \hat{\Psi}(x)\rangle,
\end{align}
with an interaction strength of $c\!=\!g/2$, where $g\!\simeq \!2 \hbar \omega_\perp a_s$ away from confinement induced resonances \cite{Olshanii1:998}. The interaction strength $c$ is experimentally controllable through, e.g., the transverse trapping frequency $\omega_\perp$, or the 3D $s$-wave scattering length, $a_s$, via a magnetic Feshbach resonance \cite{chin2010feshbach}.

The uncontrolled Hamiltonian operator of the Lieb-Liniger model, $\hat{H}_0$, corresponds to the kinetic energy, and is given by
\begin{equation}
\begin{split}
\begin{aligned}\label{eq:H_0_LL}
    \hat{H}_0 = -\frac{\hbar^2}{2m} \int_0^L dx \hat{\Psi}^\dagger(x) \frac{\partial^2}{\partial x^2} \hat{\Psi}(x).
\end{aligned}
\end{split} 
\end{equation}
Notably, the interaction-driven quantum Otto cycle for this model was recently investigated under both quasi-static \cite{chen2019interaction} and sudden quench \cite{watson2025quantum} protocols.

We highlight here that the interaction operators listed in Eq.~\eqref{eq:def} are only a small subset of all possible interaction operators to which our methods apply. Further, as the focus of this paper is to introduce the general methods of the sudden-quench Otto engine cycle, in the following we focus on continuous models with contact interactions, i.e., Eqs.~\eqref{eq:bosons} and \eqref{eq:fermions} with delta function interactions (see Section \ref{sec:LL_uniform} for further details). We note, however, that both Eqs.~\eqref{eq:BoseHubbard} and \eqref{eq:FermiHubbard} may be considered as corresponding to a discretized version of Eqs.~\eqref{eq:bosons} and \eqref{eq:fermions}, respectively, for the case of contact (i.e. on-site) interactions. Further, the case of Ising interactions expressed in Eq.~\eqref{eq:Heisenberg} was recently investigated for an Otto cycle under an infinitesimal adiabatic interaction quench of the transverse-field Ising model in Ref.~\cite{sajitha2025quantum}.

Beginning from an equilibrium quantum state, described by the initial (\textit{i}) density matrix $\hat{\rho}_i$, the corresponding equilibrium energy of the system is expressed as an expectation value $ \langle \hat{H} \rangle_i = \,\mathrm{Tr} \left[\hat{\rho}_i \hat{H}_i \right]$, where $\hat{H}_i$ denotes the total Hamiltonian prior to the quench. The work strokes of our Otto cycle, which consist of decoupling from a reservoir and suddenly quenching the interaction strength $c_i\! \to \!c_f$ over time $\delta t$, may be approximated via a zeroth-order expansion of the time evolution operator $\hat{U}(t_0 + \delta t,t_0) \equiv \hat{U}(\delta t)$ as $\hat{U}(\delta t) \simeq \hat{1}$. The benefit of such an approximation is that it allows us to estimate the work in such a unitary stroke, $W_{i\to f} \!=\!\langle\hat{H} \rangle_f \!-\! \langle\hat{H} \rangle_i$, in a particularly simple form,
\begin{equation}\label{eq:W_if}
    W_{i\to f}\simeq (c_f - c_i)  \big\langle \hat{G}_2 \big\rangle_i,
\end{equation} 
due to the fact that we are assuming the sudden quench is so rapid that the state of the system has not had time to appreciably evolve. This reduces the problem of evaluating the work of a single stroke in our Otto cycle to an equilibrium calculation of the expectation of a single Hamiltonian term, namely the expectation value $\langle \hat{G}_2\rangle_i$ in the initial thermal equilibrium state prior to the quench.

\begin{figure}[!t]
\centering
\includegraphics[width=0.45\textwidth]{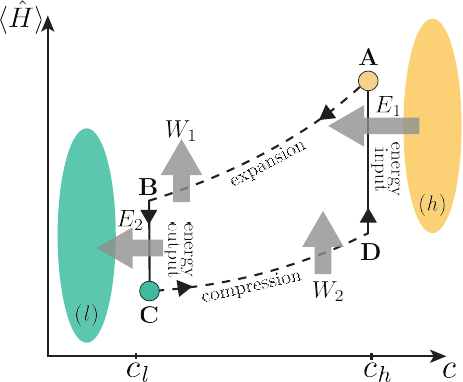}
\caption{Hamiltonian energy $\langle \hat{H} \rangle$ versus interaction strength of the working medium, for an interaction-driven quantum many-body Otto engine cycle operating between two interaction strengths $c_h$ and $c_l$ and in periodic connection to two reservoirs denoted (\textit{h}) for the high energy and (\textit{l}) for the low energy reservoir. Unitary work strokes $\mathbf{A}\!\to\!\mathbf{B}$ and $\mathbf{C}\!\to\!\mathbf{D}$ are denoted via dashed lines to signify the fact that these strokes are accomplished via a sudden quench rather than by passing through the intermediate states tracing these lines.}
\label{fig:interaction_Engine_Diagram}
\end{figure}

\subsection{Sudden quench Otto cycle}\label{sec:Otto_cycle}

The interaction-driven sudden quench Otto engine cycle consists of four strokes operating between two reservoirs denoted $(h)$ (higher energy) and $(l)$ (lower energy) in Fig.~\ref{fig:interaction_Engine_Diagram}: 
\begin{itemize}
\vspace{-0.2cm}
\item[(1)] \textit{Unitary expansion,} $\mathbf{A}\!\to\!\mathbf{B}$: the working fluid, which is initially in an equilibrium state $\hat{\rho}_h$ at an interaction strength $c_h$, is disconnected from the reservoir  ($h$) and has its interaction strength suddenly quenched from $c_h\! \to \!c_l$, with $c_h\! > \!c_l$ and energy difference $\langle \hat{H}\rangle_{\textbf{B}} \!-\!\langle \hat{H}\rangle_{\textbf{A}}\!<\!0$. This means that the work $W_1\!=\!\langle \hat{H}\rangle_{\textbf{B}} \!-\!\langle \hat{H}\rangle_{\textbf{A}}\!<\!0$ is done by the fluid. Here, $\langle \hat{H}\rangle_{\textbf{J}}$ is the expectation value of the total Hamiltonian given by Eq.~\eqref{eq:hamiltonian}, i.e., the total energy of the system, in state $\textbf{J}=\{\textbf{A,B,C,D}\}$ shown in Fig.~\ref{fig:interaction_Engine_Diagram}.
\vspace{-0.24cm}
\item[(2)] \textit{Thermalization with reservoir} (\textit{l}), $\mathbf{B}\!\to\!\mathbf{C}$: the nonequilibrium working fluid is connected to reservoir (\textit{l}) and is allowed to equilibrate while keeping the interaction strength, $c_l$, constant. The working fluid ejects energy $E_1\!=\!\langle \hat{H}\rangle_{\textbf{C}} \!-\!\langle \hat{H}\rangle_{\textbf{B}}\!<\!0$ (which can be in the form of, e.g., heat, if the contact with the reservoir is purely thermal) into the reservoir.
\vspace{-0.24cm}
\item[(3)] \textit{Unitary compression}, $\mathbf{C}\!\to\!\mathbf{D}$: the working fluid, now in an equilibrium state described by $\hat{\rho}_l$, is decoupled from reservoir (\textit{l}) and has its interaction strength suddenly quenched $c_l\! \to\!c_h$, resulting in work $W_2\!=\!\langle \hat{H}\rangle_{\textbf{D}} \!-\!\langle \hat{H}\rangle_{\textbf{C}}\!>\!0$ done on the fluid. 
\vspace{-0.24cm}
\item[(4)] \textit{Thermalization with reservoir} (\textit{h}), $\mathbf{D}\!\to\!\mathbf{A}$: the nonequilibrium working fluid is connected to reservoir (\textit{h}), allowing for energy exchange at constant $c_h$, thus taking in energy $E_2\!=\!\langle \hat{H}\rangle_{\textbf{A}} \!-\!\langle \hat{H}\rangle_{\textbf{D}}\!>\!0$ from the reservoir, and returning to its original equilibrium state $\hat{\rho}_h$. 
\end{itemize}

The net work, $W \!=\! W_1 \!+\! W_2$, of such an engine cycle, where the unitary work strokes are evaluated using Eq.~\eqref{eq:W_if}, is therefore given by
\begin{equation}\label{eq:Work_SQ}
	W \simeq - (c_h -c_l) \left(\big\langle \hat{G}_2 \big\rangle_h -  \big\langle \hat{G}_2 \big\rangle_l \right).
\end{equation}
Equation \eqref{eq:Work_SQ} means that the net work $W$ in this sudden interaction quench Otto engine cycle is determined by the \emph{correlations} of the gas in its two equilibrium states, $(h)$ and $(l)$.

Such a cycle generates net beneficial work (done by the fluid) if the total work $W \!=\! W_1 \!+\! W_2\!<\!0$, i.e., if $|W_1|>W_2$ (or $E_1>|E_2|$), with a generalised engine efficiency 
\begin{equation}
\eta = -\frac{W}{E_1} = 1 - \frac{|E_2|}{E_1}, 
\label{eq:efficiency_def}
\end{equation}
where we used the conservation of energy $W+E=0$, with $E=E_1 + E_2$ being the total energy. We note that this generalised efficiency is distinct from a typical thermodynamic efficiency, which is defined for a pure \emph{heat} engine cycle where the reservoir contact only allows for energy exchange in the form of heat. This generalised efficiency, however, extends this to scenarios where alternative forms of energy can be exchanged with the reservoir. An example of this is  chemical work (see, e.g., in Refs.~\cite{keller2020feshbach}), where the additional energy exchange can be facilitated via particle flow from or into the reservoir depending on the chemical potential imbalance.

For the interaction-driven sudden quench Otto cycle this generalised efficiency can be rewritten as
\begin{equation}\label{eq:efficiency}
    \eta \simeq 1 - \frac{\langle\hat{H}\rangle_h - \langle\hat{H}\rangle_l -  (c_h -c_l)\big\langle \hat{G}_2 \big\rangle_h}{\langle\hat{H}\rangle_h - \langle\hat{H}\rangle_l -  (c_h -c_l)\big\langle \hat{G}_2 \big\rangle_l}.
\end{equation}
Equations \eqref{eq:Work_SQ} and \eqref{eq:efficiency} represent the first key result of this work, and generalize the results first derived in Ref.~\cite{watson2025quantum} to systems with arbitrary two-body interactions, such as those given in Eq.~\eqref{eq:def}.

\subsection{Bounds on performance}\label{sec:bounds}

The above approximation for the net work and efficiency under a sudden quench protocol enables general principles to be derived for Otto engine performance.
First, for such a cycle operating under a fixed ratio of interaction strengths, $c_h/c_l$, the net work, given in Eq.~\eqref{eq:Work_SQ}, is enhanced by maximizing the difference of the total atom-atom correlations in the two equilibrium states, $\big\langle \hat{G}_2 \big\rangle_h \!-\!  \big\langle \hat{G}_2 \big\rangle_l $.
Such an engine cycle is therefore best suited to physical systems that have large difference between the values of $\big\langle \hat{G}_2 \big\rangle_h$ and $  \big\langle \hat{G}_2 \big\rangle_l$.

Second, from the definition of our Hamiltonian in Eq.~\eqref{eq:hamiltonian}, the equilibrium expectation value of the total energy separates into $\langle \hat{H} \rangle = \langle \hat{H}_0 \rangle + c \langle \hat{G}_2 \rangle$. We may therefore re-express the generalized efficiency of a sudden interaction-quench engine, given in Eq.~\eqref{eq:efficiency}, as
\begin{equation}
\begin{split}
\begin{aligned}\label{eq:efficiency_2}
    \eta \simeq 1 - \frac{\langle\hat{H_0}\rangle_h - \langle\hat{H_0}\rangle_l + c_l \left( \big\langle \hat{G}_2 \big\rangle_h - \big\langle \hat{G}_2 \big\rangle_l \right)}{\langle\hat{H_0}\rangle_h - \langle\hat{H_0}\rangle_l + c_h \left( \big\langle \hat{G}_2 \big\rangle_h - \big\langle \hat{G}_2 \big\rangle_l \right)}.
\end{aligned}
\end{split}
\end{equation}
Noting that both numerator and denominator in the fraction contain mostly the same terms, we rearrange this generalized efficiency as
\begin{equation}\label{eq:efficiency_rearranged}
    \eta \simeq 1 \!-\! \frac{c_l }{c_h }\frac{\frac{1}{c_l }\left(\langle\hat{H_0}\rangle_h \!-\! \langle\hat{H_0}\rangle_l\right) \!+\!   \big\langle \hat{G}_2 \big\rangle_h \!-\! \big\langle \hat{G}_2 \big\rangle_l}{\frac{1}{c_h }\left(\langle\hat{H_0}\rangle_h \!-\! \langle\hat{H_0}\rangle_l\right) \!+\!   \big\langle \hat{G}_2 \big\rangle_h \!-\! \big\langle \hat{G}_2 \big\rangle_l}.
\end{equation}
Recalling that for our Otto engine cycle we require $c_l/c_h<1$, we observe that the second fraction in the second term is necessarily greater than $1$. Therefore, we find that the generalised efficiency is upper-bounded by
\begin{equation}\label{eq:efficiency_bound}
    \eta < 1 -  \frac{c_l}{c_h}.
\end{equation}

We note that, while similar bounds to Eq.~\eqref{eq:efficiency_bound} have been derived previously for interaction-driven Otto engine cycles (see, e.g., Refs.~\cite{chen2019interaction,keller2020feshbach} for the 1D Bose gas), the upper bound on efficiency provided in Eq.~\eqref{eq:efficiency_bound} is entirely agnostic to the type of system-reservoir contact utilized in our Otto engine cycle.
Indeed, as this bound makes no reference to any of the other parameters (apart from the interaction strength) defining the equilibrium states that the cycle operates between, nor the type of energy that flows between the reservoirs and the working fluid, this formula provides a universal upper bound on \emph{all} sudden quench cycles operating between the same ratio of interaction strengths, for all types of models specified in Eq.~\eqref{eq:def}, and for all forms of system-reservoir contact.

Lastly, for a quasi-static Otto engine cycle, in the limit of an infinitesimal quench, $c_h -c_l \equiv \delta c$ for $\delta c \!\to\! 0$, we may approximate the evolved state by expanding the energy difference in a Taylor series \cite{sajitha2025quantum},
\begin{equation}
	\langle \hat{H} \rangle_f - \langle \hat{H} \rangle_i \simeq   \bigg\langle \frac{d \hat{H}}{dc} \bigg\rangle_{\!\!i}  \delta c + \dots \, .
\end{equation}
The net work of this infinitesimal quasi-static quench may therefore be approximated by
\begin{equation}\label{eq:Work_Ad}
	W^{qs} \simeq - \delta c \left(\bigg\langle \frac{d \hat{H}}{dc} \bigg\rangle_{\!\!h} - \bigg\langle \frac{d \hat{H}}{dc} \bigg\rangle_{\!\!l}  \right).
\end{equation}
We then note that these expectation values corresponds exactly to a typical application of the Hellmann-Feynman theorem \cite{hellmann1933,feynman1939forces}, i.e. $\langle d\hat{H} / dc\rangle = \langle \hat{G}_2\rangle$, meaning
\begin{equation}
    W^{qs} \simeq - \delta c \left(\big\langle \hat{G}_2 \big\rangle_h -  \big\langle \hat{G}_2 \big\rangle_l \right),
\end{equation}
which corresponds exactly to that for a sudden quench, given in Eq.~\eqref{eq:Work_SQ}. Likewise, the efficiency of this quasi-static Otto cycle under an infinitesimal quench takes the form
\begin{equation}\label{eq:efficiency_qs}
    \eta^{qs} \simeq 1 - \frac{\langle\hat{H}\rangle_h - \langle\hat{H}\rangle_l -  \delta c\big\langle \hat{G}_2 \big\rangle_h}{\langle\hat{H}\rangle_h - \langle\hat{H}\rangle_l -  \delta c\big\langle \hat{G}_2 \big\rangle_l}.
\end{equation}
As the efficiency of this infinitesimal quasi-static quench takes the same form as that of the sudden quench, given in Eq.~\eqref{eq:efficiency}, it must also conform to the upper bound expressed in Eq.~\eqref{eq:efficiency_bound}.

\section{Application of the interaction-driven Otto engine }\label{sec:applications_interaction}

In this section, we demonstrate applications of the interaction-driven quantum Otto engine cycle introduced in Section.~\ref{sec:Otto}. 
In particular, we first utilize the uniform Lieb-Liniger model, whose sudden-quench quantum Otto engine cycle was previously investigated in Ref.~\cite{watson2025quantum}, to illustrate the various bounds on engine performance derived in Section.~\ref{sec:bounds}.
We then provide further examples of the sudden quench Otto engine for other experimentally relevant physical models, including a harmonically trapped ultra-cold Bose gases where the role of system dimension is investigated, and in the Yang-Gaudin model of spin-$1/2$ fermions in 1D, where we introduce a novel quantum Otto engine cycle in which control over the polarization of the working fluid is shown to greatly enhance engine operation.

\begin{figure}[!t]
\centering
\includegraphics[width=0.46\textwidth]{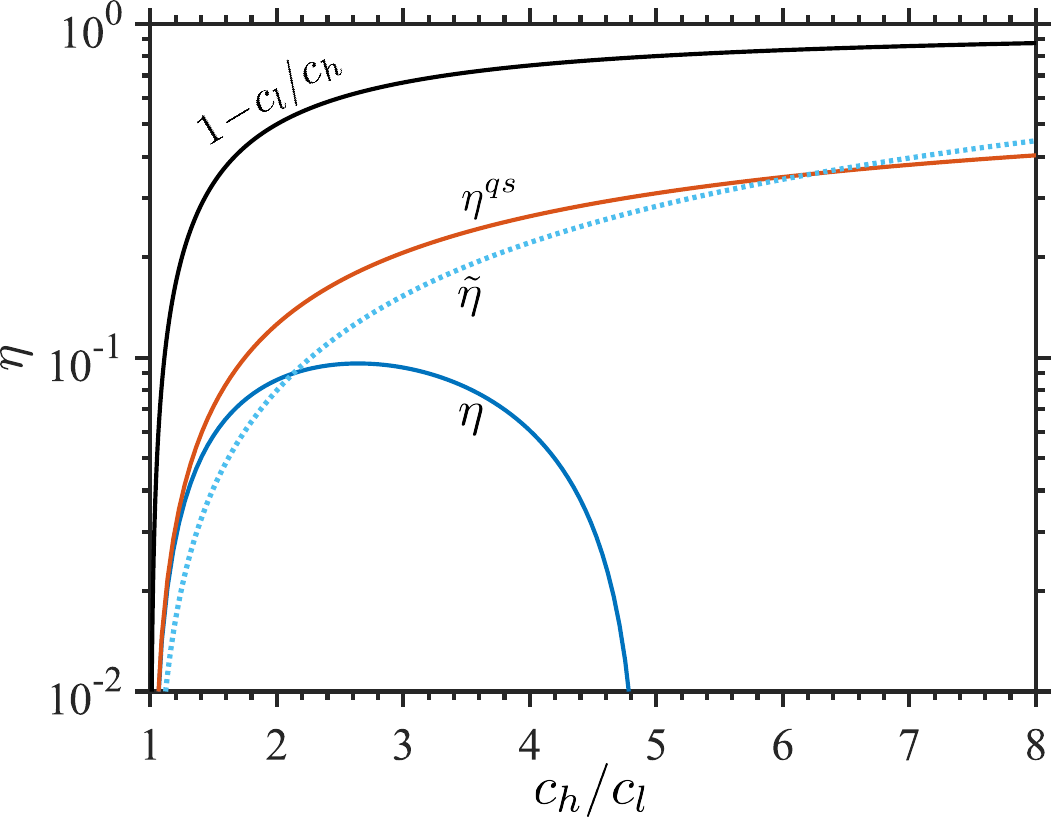}
\caption{Efficiency of the interaction-driven sudden quench Otto engine cycle for a uniform 1D Bose gas in the low temperature, weakly interacting regime (see text), operating between two reservoirs at a fixed ratio of temperatures, $T_h/T_l\!=\!3$. 
The case of purely thermal reservoir contact is demonstrated for the sudden quench, $\eta$ (solid blue line), and quasi-static, $\eta^{qs}$ (solid red line), protocols. We observe that these two efficiencies coincide under small interaction strength ratios, where $c_h\!-\!c_l \!\to\!0$.
This may be contrasted with the efficiency of the thermochemical engine cycle, $\tilde{\eta}$, which includes chemical work accomplished via diffusive contact with the reservoirs; in this example the density difference between the hot, $\rho_h$, and cold, $\rho_l$, equilibrium states of the system is chosen to be equal to $\Delta \rho/\rho_h \!=\!0.2$. All efficiencies are compared with the absolute upper bound derived in the text, given by $1\!-\!c_l/c_h$.}
\label{fig:LiebLiniger_efficiency}
\end{figure}

\subsection{Lieb-Liniger gas}\label{sec:LL_uniform}

We begin with an illustrative example of the bounds on engine performance derived in Section.~\ref{sec:bounds}. In particular, we examine the efficiency of the interaction-driven Otto engine cycle where the working fluid is a 1D Bose gas with contact interactions. This physical system corresponds to the Lieb-Liniger model, whose Hamiltonian is expressed in Eqs.~\eqref{eq:g2_LL} and \eqref{eq:H_0_LL}, and for which operation as a heat engine (as well as an accelerator, heater, and refrigerator) was recently studied in Ref.~\cite{watson2025quantum}. For simplicity, we focus here on the regime of weak interactions, $\gamma \!\ll\!1$, where $\gamma \!=\!m g /\hbar^2 \rho$ is the dimensionless interaction strength for a uniform system of 1D density $\rho\!=\!N/L$, in the low temperature quasicondensate regime, $2\gamma \!\ll\! \tau \!\ll\!2 \sqrt{\gamma}$, where $\tau \!=\! T/T_d$ is the dimensionless temperature expressed in terms of the temperature of quantum degeneracy, $T_d \! =\! \hbar^2 \rho^2 / 2 m k_B$.  Operation outside the chosen region was investigated in Ref.~\cite{watson2025quantum}, alongside performance under quasistatic interaction driving, which is utilized here for the purposes of illustrating the bounds on performance.

The efficiency of the sudden interaction-quench Otto engine with the Lieb-Liniger 1D Bose gas as the working fluid was shown in  Ref.~\cite{watson2025quantum} to be of the form given in Eq.~\eqref{eq:efficiency}, with $\langle \hat{G}_2 \rangle \!=\!N \rho g^{(2)}(0)$ for a translationally invariant (uniform) system. In detail, $g^{(2)}(0)$ is the local (same-point) normalized atom-atom correlation function, defined by
\begin{equation}
    g^{(2)}(x,x) = \frac{  \langle \hat{\Psi}^\dagger(x) \hat{\Psi}^\dagger(x) \hat{\Psi}(x) \hat{\Psi}(x \rangle}{\rho(x)^2},
    \label{normalized_g2}
\end{equation}
also known as Glauber's local second-order correlation function \cite{glauber1963,glauber2006}.
The correlation function $g^{(2)}(0)$ can be evaluated analytically in six asymptotic regimes of the uniform 1D Bose gas, expressed in terms of $\gamma$ and $\tau$ \cite{kheruntsyan2003pair,kheruntsyan2005finite,kerr2023analytic}, as well as exactly for any $\gamma$ and $\tau$ using the thermodynamic Bethe ansatz \cite{yang1969thermodynamics}.

In the low temperature, weakly interacting regime ($2\gamma \!\ll\! \tau \!\ll\!2 \sqrt{\gamma}$, $\gamma\ll 1$), there exist well-known analytic approximations for both the correlation function \cite{kheruntsyan2003pair,kerr2023analytic},
\begin{equation}
    g^{(2)}(0) \simeq 1 + \frac{\tau}{2\sqrt{\gamma}},
\end{equation}
and the total energy of the uniform system at finite temperature \cite{kerr2023analytic},
\begin{equation}
    \langle \hat{H} \rangle \!\simeq\! N\frac{\hbar^2 \rho^2}{2m}\left(\gamma + \frac{\zeta(3/2)}{4\sqrt{\pi}}\tau^{3/2} + \frac{\zeta(1/2)}{2\sqrt{\pi}}\tau^{1/2}\gamma \right),
\end{equation}
where $\zeta(n)$ is the Riemann zeta function for some rational number $n$. These analytic approximations allow us to evaluate the efficiency of the interaction-driven Otto heat engine cycle for a sudden quench, which is shown as the solid blue curve labeled $\eta$ in Fig.~\ref{fig:LiebLiniger_efficiency}, which was demonstrated to be in excellent agreement with exact calculations using the thermodynamic Bethe ansatz in Ref.~\cite{watson2025quantum}. 
We note that engine operation occurs over a finite range of the interaction strength ratio for the sudden quench Otto heat engine. This occurs due to the dependence of the interatomic correlations on the interaction strength, and results in inevitable operation as a `heater' (instead of an `engine') beyond a sufficiently large critical value of $c_h/c_l$ (see Ref.~\cite{watson2025quantum} for further details).

Further, the known analytic formula for entropy in the same regime  \cite{kerr2023analytic},
\begin{equation}
    S  \!\simeq\!  N k_B\left( \frac{3 \zeta(3/2)}{4 \sqrt{\pi}} \sqrt{\tau} - \sqrt{\gamma} - \frac{ \zeta(1/2)}{2 \sqrt{\pi}} \frac{\gamma}{\sqrt{\tau}} \right),
\end{equation}
allows us to evaluate the efficiency of the unitary quasi-static engine cycle, where the work strokes preserve $S$. Quasi-static efficiency is shown in Fig.~\ref{fig:LiebLiniger_efficiency} as the solid red line, and denoted $\eta^{qs}$. We observe that the sudden and quasi-static Otto engine cycles converge in the limit $c_h/c_l \!\to \! 1$ (i.e. $c_h\!-\!c_l \!\to\! 0$), as explained in Sec.~\ref{sec:bounds}.

Above, we considered cycles where \emph{only} heat is exchanged with the two external reservoirs (i.e. thermal contact). However, we may consider instead the situation where diffusive contact \emph{in addition} to thermal contact is allowed, which may be accomplished via an intake of particles from the high-energy reservoir ($h$) in addition to heat. Provided the same total number of particles are removed from the working fluid during contact with the low-energy reservoir ($l$), such a system may be considered to operate as a thermochemical engine \cite{Nautiyal_2024}.

The efficiency of the thermochemical Otto engine cycle is shown as the blue dotted line in Fig.~\ref{fig:LiebLiniger_efficiency}, denoted $\tilde{\eta}$, calculated using the same method as the sudden quench heat engine efficiency, $\eta$, given in Eq.~\eqref{eq:efficiency}. In particular, during the equilibration strokes ($\mathbf{B\to C}$ and $\mathbf{D\to A}$ in Fig.~\ref{fig:interaction_Engine_Diagram}) $\Delta N$ particles are exchanged with the reservoir such that the density difference, $\Delta \rho \!=\!\rho_h \!-\! \rho_l$, between the equilibrium states $\mathbf{A}$ and $\mathbf{C}$ in Fig.~\ref{fig:interaction_Engine_Diagram} satisfies $\Delta \rho/\rho_h \!=\!0.2$. Such a generalized efficiency is no longer limited by the typical bounds on engine performance given by $\eta^{qs}$, as chemical work may, in principle, be entirely converted into mechanical work.
Therefore, an additional exchange of particles during system-reservoir contact renders such a nonequilibrium thermochemical Otto engine capable of exceeding the adiabatic limit of the Otto engine where the system-reservoir contact consists of exclusively heat.
However, we see that all sudden-quench engine cycles still remain bounded by $1\!-\!c_l/c_h$, derived in Section.~\ref{sec:bounds} for arbitrary reservoir contact.

\subsection{Harmonically trapped Bose gas}\label{sec:bose_gas_interaction}

Next, we examine operation of the interaction-driven Otto cycle in harmonically trapped Bose gases in dimensions $d\!=\!1,2,3$.
The uncontrolled operator terms consist of kinetic energy and external trapping potential energy,
\begin{align}
	\hat{H}_0	=  -&\frac{\hbar^{2}}{2m}  \int d\mathbf{r}\,  \hat{\Psi}^{\dagger}(\mathbf{r}) \,\nabla_{\mathbf{r}}^2 \hat{\Psi}(\mathbf{r}) \\ 
    &+ \int  d\mathbf{r}\, V(\mathbf{r})  \hat{\Psi}^{\dagger}(\mathbf{r}) \hat{\Psi}(\mathbf{r}),
\end{align}
where $\mathbf{r}$ denotes atomic position in $d$ dimensions, and the gas is contained in a static and isotropic harmonic trap of frequency $\omega$, $V(\mathbf{r})\!=\!m \omega^2 \mathbf{r}^2/2$. We note here that the interaction-driven Otto engine in harmonically trapped Bose gases have been previously investigated theoretically in Refs.~\cite{li2018efficient,keller2020feshbach,Nautiyal_2024}, and recently realised experimentally in Refs.~\cite{koch2022making,simmons2023thermodynamic}.

Control over interparticle interaction strength allows us to express the net work in terms of the integrated local second-order correlation function, given in Eq.~\eqref{eq:g2_LL}, as was the case for the uniform 1D system. However, for the harmonically trapped gas, this integration must be performed over the inhomogeneous profile of the correlation function. To do so, we utilize the \emph{normalized} local second-order correlation function, similar to the one introduced for the uniform 1D system in Eq.~\eqref{eq:g2_LL}.
Rearranging and integrating over $\mathbf{r}$, we express our total integrated correlation as,
\begin{equation}
    \big\langle \hat{G}_2 \big\rangle = \int d\mathbf{r} g^{(2)}(\mathbf{r},\mathbf{r}) \rho(\mathbf{r})^2.
\end{equation}

To continue, we assume that our system is in the low temperature, weakly interacting regime, such that it can be modeled using the Thomas-Fermi (inverted parabola) approximation for the density profile, $\rho(\mathbf{r})$, with a coherent-state correlation of $g^{(2)}(\mathbf{r},\mathbf{r}) \!\simeq\! 1$ \cite{pitaevskii2016bose,pethick2008bose}.
Combining this approximation with Eq.~\eqref{eq:Work_SQ}, and supplying the $h$ (high-energy) and $l$ (low-energy) subscripts to the respective equilibrium state quantities, we arrive at the following result for the net work of this sudden interaction quench Otto cycle in $d$ dimensions:
\begin{equation}\label{eq:W_sq_g}
	\frac{W}{\hbar \omega} \!\simeq\! - \frac{2}{(d+4)}  (\overline{g}_h \!-\! \overline{g}_l) \Big( \overline{\rho}_h(0) N_h \!-\! \overline{\rho}_l(0) N_l  \Big).
\end{equation}
Here, we have additionally converted to dimensionless quantities defined in natural harmonic oscillator units, so that the dimensionless interaction strength is $\overline{g}_{h(l)}\!=\!g_{h(l)}/(\hbar \omega l_{\mathrm{ho}})$, where $l_{\mathrm{ho}}\!=\!\sqrt{\hbar/m \omega}$ is the harmonic oscillator length for a harmonic trap of frequency $\omega$, $\overline{\rho}_{h(l)}(0)\!=\!\rho_{h(l)}(0)\, l_{\mathrm{ho}}$ is the dimensionless peak density at the trap center, and $N_{h(l)}$ is the total atom number of the working fluid at the respective equilibrium points, shown in Fig.~\ref{fig:interaction_Engine_Diagram}. 

As highlighted recently in Ref.~\cite{Nautiyal_2024}, operation of such an interaction-driven quantum Otto cycle as a pure heat engine, i.e., for system-reservoir contact that is only thermal, is not feasible for a harmonically trapped 1D Bose gas. In order to extract beneficial net work from such an Otto cycle, one requires particle exchange with the reservoirs in addition to, or in replacement, heat, meaning operation occurs as a thermochemical engine cycle. The reason for this requirement is made clear by inspection of Eq.~\eqref{eq:W_sq_g} where, if $N_h\!=\!N_l$, beneficial net work extraction requires $\bar{\rho}_h(0) \!>\!\bar{\rho}_l(0)$, which is difficult to satisfy when the temperature of state $h$ is higher than that of $l$, as the peak density is reduced at higher temperature due to thermal broadening of the density profile.

\begin{figure}[!t]
\centering
\includegraphics[width=0.46\textwidth]{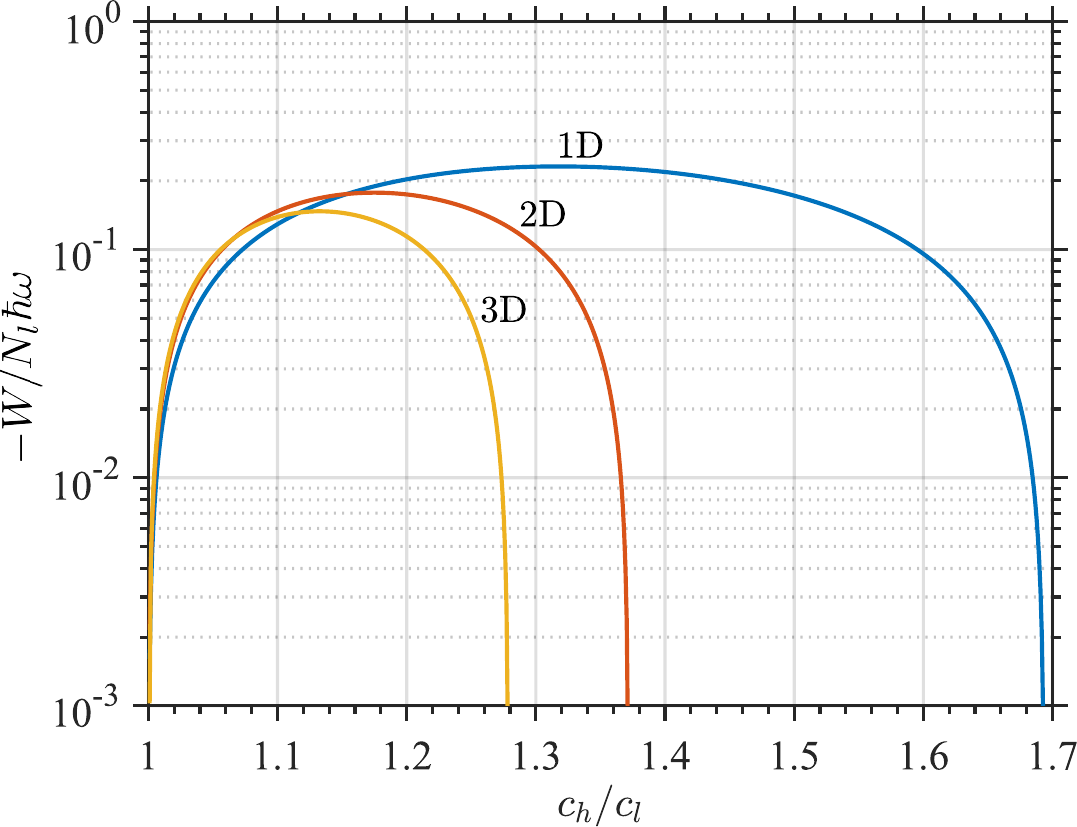}
\caption{Net work of a sudden interaction-quench Otto engine fueled by chemical work in an ultra-cold harmonically trapped Bose gas of dimensions $d\!=\!1$, $2$, and $3$ ($1D$, $2D$ and $3D$) in its $T\!=\!0$ ground state. Net work is given in natural units of the harmonic oscillator energy $\hbar \omega$, and normalized to the total particle number of the working fluid in the low energy equilibrium state $l$, given by $N_l$. This particle number is chosen such that the Thomas-Fermi approximation is valid for each realization. In particular, we have $N_h\!=\!(2\!\times\! 10^3, \,5\!\times\! 10^4, \, 1\!\times\! 10^6)$, for dimensions $d=(1,2,3)$, respectively. The low energy equilibrium state ($l$) is characterized by a dimensionless interaction strength of $\overline{g}_l\!=\!0.1$ (in harmonic oscillator units, see text), and the number of particles exchanged with the reservoirs is fixed by the ratio $\Delta N/N_l \!=\! 0.1$ in all dimensions.   }\label{fig:GP_g}
\end{figure}

For this reason, and for simplicity, in the following we consider a chemical Otto engine cycle at $T\!=\!0$, fueled solely through intake and output of particles into the $h$ and $l$ reservoirs, respectively (and hence making it a purely chemical engine, as was previously investigated for the quasistatic Otto engine cycle in Ref.~\cite{keller2020feshbach}).
The net work, given in Eq.~\eqref{eq:W_sq_g}, of the interaction-driven \emph{chemical} Otto engine cycle is shown in Fig.~\ref{fig:GP_g} for dimensions $d=1$, $2$, and $3$. As this net work utilizes the Thomas-Fermi approximation, higher dimensions require increasing total atom number for this approximation to remain valid \cite{pitaevskii2016bose,pethick2008bose}. Hence, to compare net work across physical systems with differing dimension, we additionally normalize the net work to the total particle number in equilibrium state $l$, $N_l$, and fix the number of particles exchanged with the reservoir, $\Delta N$, to be proportional, $\Delta N/N_l \!=\!0.1$.

We note that both the maximum height and breadth over which this cycle operates diminishes with increasing dimension. Upon inspection of the formula for net work given in Eq.~\eqref{eq:W_sq_g}, we observe that the extraction of beneficial net work, i.e. $W\!<\!0$, corresponds to the region where $\overline{\rho}_h(0) N_h \!>\! \overline{\rho}_l(0) N_l$, since $\overline{g}_h \!>\!\overline{g}_l$ by definition. 
Thus, the engine performance is entirely dependent on the scaling of the peak density with dimension $d$,
\begin{equation}
    \bar{\rho}(0) = \rho(0)l_{ho}= \left(\frac{d(d+2)N}{2\Omega_d} \right)^{\frac{2}{d+2}}\left(\frac{1}{2\overline{g}}\right)^{\frac{d}{d+2}},
\end{equation}
where $\Omega_d$ is an angular geometric factor in dimension $d$, with $\Omega_1\!=\!2$, $\Omega_2 \!=\!2\pi$, and $\Omega_3 \!=\!4 \pi$. The inverse scaling of peak density with the interaction strength clearly degrades with increasing dimension, which results in a decrease of the range over which the engine cycle obtains beneficial net work.

\subsection{Yang-Gaudin model}

The Yang-Gaudin model is a paradigmatic and experimentally realizable fermionic model, describing a uniform spin-1/2 1D Fermi gas with contact interactions between the two spin components \cite{GAUDIN196755,yang2967someexact,guan2013fermi,liao2010spin,moritz2005confinement,truscott2001fermi,partridge2005molecular}. The Hamiltonian of this model consists of the kinetic energy term,
\begin{align}
	&\hat{H}	_0= -\sum_{\sigma=\uparrow,\downarrow}\frac{\hbar^{2}}{2m}  \int_{0}^{L}  dx\,  \hat{\Psi}_{\sigma}^{\dagger}(x) \frac{\partial^2 \hat{\Psi}_{\sigma}(x)}{\partial x^2}, 
\end{align}
where $\hat{\Psi}^\dagger_{\uparrow(\downarrow)}(x)$ is the field creation operator for the spin up (down) component, and a contact interaction term between the two spin components,
\begin{equation}
    	\hat{H}_{\text{int}}=c\,\hat{G}_2=  c\int_{0}^{L}  dx\, \hat{\Psi}_{\uparrow}^{\dagger}(x) \hat{\Psi}_{\downarrow}^{\dagger}(x) \hat{\Psi}_{\downarrow}(x) \hat{\Psi}_{\uparrow}(x).
\end{equation}
Here, $c\!=\!g_{1D}$ is the interaction strength, where $g_{1D}\!\simeq\!2 \hbar \omega_\perp a_s$ when realized experimentally via strong transverse harmonic confinement of frequency $\omega_{\perp}$ \cite{Olshanii1:998,Bergeman2003confinement,guan2013fermi}, and $a_s$ is the 3D $s$-wave scattering length between the two spin components.
Importantly, as the interactions are only between the opposite spin components, the correlation may be enhanced or suppressed depending on the polarization $\mathcal{P}=(n_{\uparrow} - n_{\downarrow})/n$, where $n=n_{\uparrow}+n_{\downarrow}$ is the total uniform density, and $n_{\uparrow(\downarrow)}=\langle \hat{\Psi}^{\dagger}_{\uparrow(\downarrow)}(x)\hat{\Psi}_{\uparrow(\downarrow)}(x) \rangle $ is the density in the spin up (down) component.

At sufficiently high temperatures, $k_B T \gg \hbar^2 c^2/2m$, one may express the expectation value of the total atom-atom correlation as
\cite{patu2016thermodynamics,he2016universal},
\begin{align}
\label{eq:g2_yanggaudin}
	&\langle \hat{G}_2\rangle =
    L\langle \hat{\Psi}_{\uparrow}^{\dagger}(x) \hat{\Psi}_{\downarrow}^{\dagger}(x) \hat{\Psi}_{\downarrow}(x) \hat{\Psi}_{\uparrow}(x)\rangle \nonumber\\
    &= L \frac{n^2 (1 \!-\! \mathcal{P}^2)}{4}  
    \left( 1 \!-\! \frac{\sqrt{\pi}  \gamma}{\sqrt{2 \tau}} e^{\frac{ \gamma^2}{2\tau}} \left[ 1 \!-\! \mathrm{erf}\left( \frac{\gamma}{\sqrt{2\tau}}\right) \right]\right),
\end{align}
where we have introduced a dimensionless interaction strength $\gamma\!=\!2cm/\hbar^2n$  for constant total density $n$, and a dimensionless temperature $\tau\!=\!2 m k_B T / \hbar^2 n^2$. We may utilize this analytic expression for $\langle \hat{G}_2\rangle$ to evaluate the performance, in particular the net work given in Eq.~\eqref{eq:Work_SQ}, of the interaction-driven sudden quench Otto engine cycle---in which the working fluid is a spin-1/2 Fermi gas in 1D---via the methods introduced in Section \ref{sec:Otto}.

\begin{figure}[!t]
\centering
\includegraphics[width=0.46\textwidth]{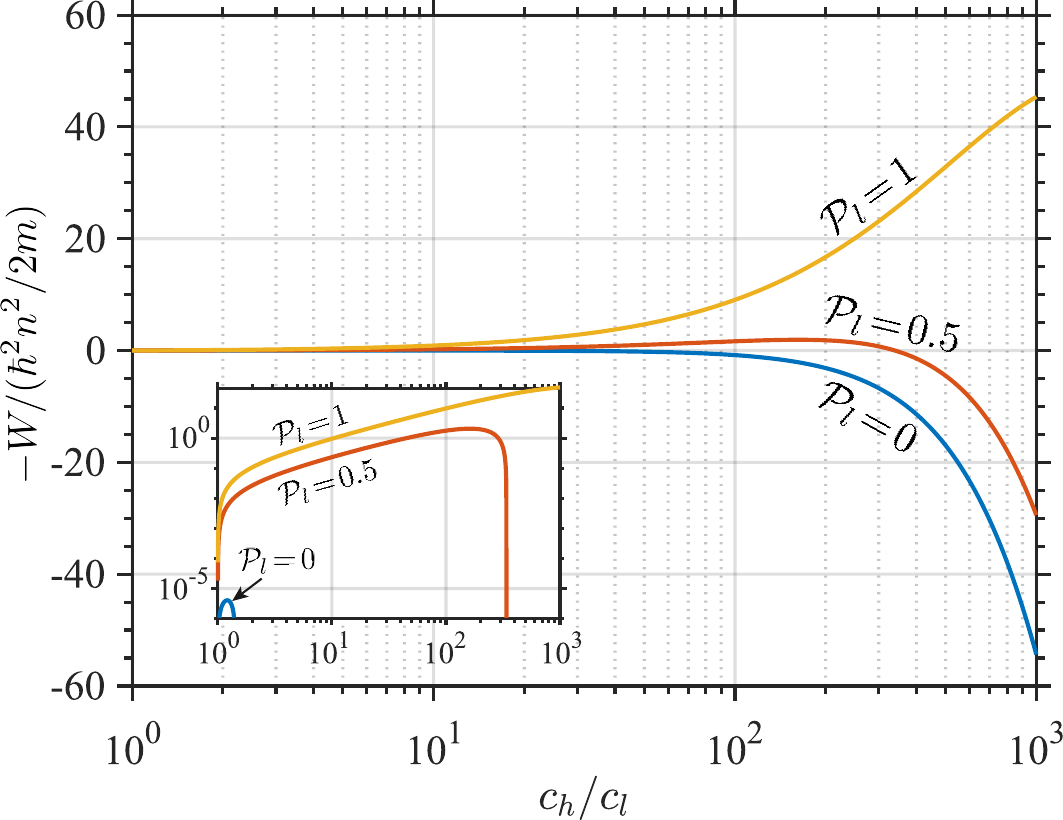}
\caption{Net work of the interaction-driven Otto engine realized in the Yang-Gaudin model, where beneficial net work is enhanced via control over spin polarization, $\mathcal{P}$. Engine operation occurs at a fixed total atom number, $N\!=\!10$, between a low energy equilibrium state defined by a temperature $\tau_l \!=\!100$ and dimensionless interaction strength $\gamma_l\!=\!0.01$ (see text), and a high energy state with temperature $\tau_h\!=\!200$. For spin polarization of $\mathcal{P}_h\!=\!0$ (solid blue line), the net work extracted and the range of $c_h/c_l$, over which the system operates as an engine ($W<0$), are very small though. However, increasing the difference in spin polarization between the high and low energy equilibrium states via control over $\mathcal{P}_l$ enhances the production of beneficial net work.
}\label{fig:Yang_Gaudin}
\end{figure}

As the net work depends directly on the correlations present in the thermal equilibrium states, we may utilize the dependence of $\langle \hat{G}_2\rangle$ on the polarization, $\mathcal{P}$, to examine its effect on engine performance.
In particular, we begin by fixing the polarization in the high energy state, $\mathcal{P}_h\!=\!0$, in order to maximize the total correlation, which scales as $\langle \hat{G}_2\rangle\propto (1-\mathcal{P}^2)$, as shown in Eq.~\eqref{eq:g2_yanggaudin}.
Then, given a fixed temperature ratio of the high and low energy equilibrium states, $\tau_h/\tau_l \!=\!2$, we investigate the dependence of net work on the interaction strength ratio, $c_h/c_l$, for a fixed value of polarization $\mathcal{P}_l\!=\!\mathcal{P}_h\!=\!0$. This net work is shown as the solid blue line, denoted $\mathcal{P}_l\!=\!0$, in Fig.~\ref{fig:Yang_Gaudin}. For such a cycle, we observe that beneficial net work (i.e. $W\!<\!0$) only occurs over a small span of the interaction strength ratio (see inset).

To explain this, we first re-write the total correlation $\langle \hat{G}_2\rangle$ in terms of a new dimensionless parameter $\chi \!\equiv\! \gamma^2 / 2 \tau$, which, in the high temperature regime, must satisfy $\chi \! \ll \! 1$,
\begin{equation}\label{eq:Work_YG}
	\langle \hat{G}_2\rangle  \!=\! L\frac{n^2 (1 \!-\! \mathcal{P}^2)}{4}  \bigg\{ 1 \!-\! \sqrt{\pi \chi} e^{\chi} \left[ 1 - \mathrm{erf}\left(\sqrt{\chi}\right) \right]\bigg\}.
\end{equation}
Importantly, this is a monotonically decreasing function of $\chi$ for the region of interest, $0\!<\!\chi \!\ll\!1$. Hence, as the net work of the sudden interaction quench Otto cycle, given in Eq.~\eqref{eq:Work_SQ}, requires $\langle \hat{G}_2 \rangle_h \!>\!\langle \hat{G}_2 \rangle_l$ to obtain beneficial net work ($W\!<\!0$), 
we require that $\chi_h \!<\! \chi_l$. Thus, one must satisfy the inequality $ \gamma_h^2 / 2 \tau_h \! < \! \gamma_l^2 / 2 \tau_l$. Rearranging, we find a simple criterion for extractable net work to be $ \gamma_h/\gamma_l \! < \sqrt{ \!\tau_h/\tau_l} $, meaning engine operation is restricted to $ \gamma_h/\gamma_l\!<\!\sqrt{2}$ for our fixed temperature ratio $\tau_h/\tau_l\!=\!2$ and constant polarization $\mathcal{P}_l\!=\!\mathcal{P}_h\!=\!0$. As the region of beneficial net work extraction for this cycle is small in both width and amplitude, we plot $-W$ on a logarithmic scale in the inset of Fig.~\ref{fig:Yang_Gaudin}.

Since the interparticle interactions are exclusively between the two opposite spin components, one may exercise control over the population of each spin component to enhance the performance of this Otto engine cycle. Such a scenario may experimentally consist of control over the individual spin components during the thermalization strokes, via, e.g., an rf pulse. Alternatively, one may consider the case where the high energy reservoir consists of a spin-balanced, $\mathcal{P}\!=\!0$ system, and the low energy reservoir consisting of a spin polarized, $\mathcal{P}=\!\pm1$ system. Connection to these reservoirs during the relevant thermalization strokes would result in the equilibrium states of the working fluid satisfying $|\mathcal{P}_l| >|\mathcal{P}_h|$, therefore enhancing the net extraction of work. This enhancement is demonstrated in Fig.~\ref{fig:Yang_Gaudin}, where operation between $\mathcal{P}_h \!=\!0$ and $\mathcal{P}_l\!=\!0.5$ is shown as the red solid line, and between $\mathcal{P}_h \!=\!0$ and $\mathcal{P}_l\!=\!1$ as the yellow solid line. Through this, we observe that one may exploit control over the polarization of the working fluid in order to enhance the extraction of work from the Otto engine cycle that uses a two-component Fermi gas as a working fluid.

\section{General sudden quench Otto cycles}\label{sec:generalise_otto}

In Sec. \ref{sec:Otto}, we demonstrated that the performance of an interaction-driven quantum Otto engine cycle was entirely determined by the two-body inter-particle correlations when the unitary work strokes consisted of a sudden quench of the interaction strength $c$. However, the methods employed to determine the net work and efficiency of such a sudden quench engine are entirely independent of the operator controlled during these work strokes.
Therefore, in this section we demonstrate how the methods introduced in Section \ref{sec:Otto} can be generalized to situations where the externally controllable parameters are related to arbitrary Hamiltonian operators. We demonstrate the application to the case of one-body operator corresponding to an external trapping potential. Yet, we emphasize that such a protocol is entirely general and may be applied to arbitrary (not necessarily one-body) operators.

We again introduce a Hamiltonian consisting of some specified set of uncontrolled operators, again denoted $\hat{H}_0$, and an externally-controllable Hamiltonian operator, $c \hat{\mathcal{V}}$,
\begin{equation}\label{eq:hamiltonian_multiparameter}
    \hat{H} = \hat{H}_0 +  c \, \hat{\mathcal{V}}.
\end{equation}
Operation of this sudden quench Otto cycle follows the exact protocol detailed in Section \ref{sec:Otto_cycle}, with control over interaction strength replaced by the strength, $c$, of the operator $\hat{\mathcal{V}}$, which is again quenched  between $c_l$ and $c_h$. Therefore, all formulas found in Sec.~\ref{sec:Otto} for net work, efficiency, and the various bounds on operation, apply to this case with the operator $\hat{\mathcal{V}}$ replacing $\hat{G_2}$.

\subsection{Harmonic trap quench in an ultracold Bose gas}\label{sec:generalized_application}

\begin{figure}[!t]
\centering
\includegraphics[width=0.46\textwidth]{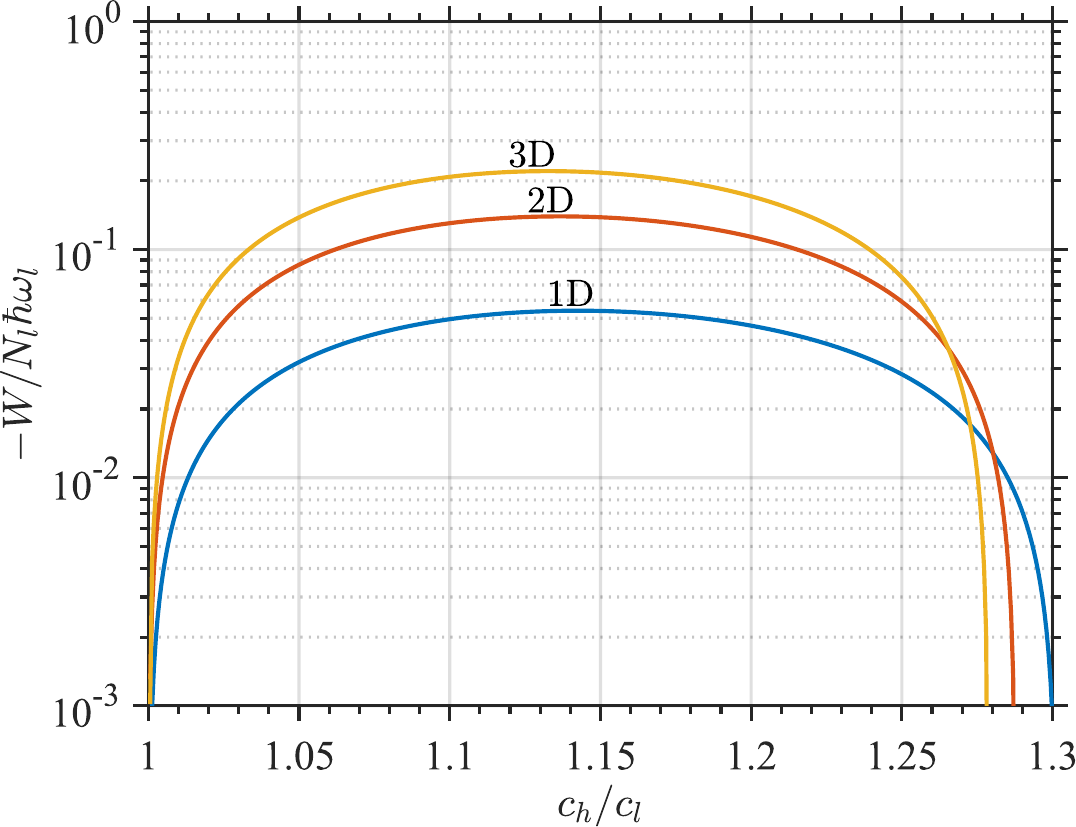}
\caption{Volumetric quantum Otto engine cycle fueled by chemical work and driven by a sudden quench of harmonic trapping frequency, with $c_{h(l)}\!=\!\omega_{h(l)}^2$, in a $d$-dimensional ultracold Bose gas with contact interactions. Parameters are chosen for the low energy thermal state $(l)$ to exactly match that used in Fig.~\ref{fig:GP_g}, with the same number of atoms exchanged between the system and reservoir during equilibration strokes.
}\label{fig:GP_omega}
\end{figure}

To illustrate the application of this sudden quench quantum Otto engine cycle, we focus on the case of a harmonically trapped ultra-cold Bose gas with contact interactions in $d=1$ to $3$ dimensions, with uncontrolled terms
\begin{align}\label{eq:LL_ho}
	\hat{H}_0	=  -&\frac{\hbar^{2}}{2m}  \int d\mathbf{r}\,  \hat{\Psi}^{\dagger}(\mathbf{r}) \,\nabla_{\mathbf{r}}^2 \hat{\Psi}(\mathbf{r}) \nonumber \\ 
    &+ \frac{g}{2} \int  d\mathbf{r}\,   \hat{\Psi}^{\dagger}(\mathbf{r}) \hat{\Psi}^{\dagger}(\mathbf{r}) \hat{\Psi}(\mathbf{r})\hat{\Psi}(\mathbf{r}),
\end{align}
where the interaction strength, $g$, is kept constant.
An external harmonic trap may be incorporated into the total Hamiltonian \eqref{eq:hamiltonian_multiparameter} via defining
\begin{equation}
        c\hat{\mathcal{V}}\equiv \hat{V} = \frac{1}{2}m\omega^2\int d\mathbf{r} \,  \mathbf{r}^2 \hat{\rho}(\mathbf{r}).
\end{equation}
with $\hat{\mathcal{V}}\equiv\frac{1}{2}m\int d\mathbf{r} \,  \mathbf{r}^2 \hat{\rho}(\mathbf{r})$, $c\!=\! \omega^2 $ corresponding to the harmonic trapping frequency squared, and $\hat{\rho}(\mathbf{r}) \!=\!\hat{\Psi}^\dagger(\mathbf{r})\hat{\Psi}(\mathbf{r})$ being the density operator.

The net work may be evaluated in terms of the equilibrium expectation value of this operator,
\begin{equation}\label{eq:Work_Ad}
	 \langle \hat{V} \rangle = \frac{1}{2} m \omega^2 \int d\mathbf{r} \rho(\mathbf{r}) \mathbf{r}^2 \equiv \frac{1}{2} m \omega^2 \langle \mathbf{r}^2 \rangle,
\end{equation}
where we have defined the atomic position variance $\langle \mathbf{r}^2\rangle$.
From this, we arrive at the following expression for net work
\begin{equation}
	W  = -\frac{1}{2} m (\omega_h^2 - \omega_l^2) \left( \langle \mathbf{r}^2 \rangle_h - \langle \mathbf{r}^2 \rangle_l \right).
\end{equation}

In the weakly interacting regime, and at sufficiently low temperatures,
the expectation value of the position variance, $\langle \mathbf{r}^2 \rangle$, can again be evaluated using the Thomas-Fermi approximation, similar to that shown in Section \ref{sec:bose_gas_interaction},
\begin{align}\label{eq:pos_variance}
   \langle \mathbf{r}^2 \rangle & =   \frac{N d}{(d+4)}  R^{2} 
\end{align}
where $R^2 = 2 g \rho(0)/(m \omega^2)$ is the squared Thomas-Fermi radius in a $d$-dimensional system.

Operation of this volumetric Otto engine cycle is demonstrated in Fig.~\ref{fig:GP_omega}, for ultracold Bose gases in dimensions $d=1$ to $3$. In particular, we investigate operation of the sudden quench Otto cycle which is fueled by chemical work via particle transfer with the high and low energy reservoirs. The parameters defining the low energy equilibrium state $l$, and the number of particles transferred to and from the working fluid during system-reservoir contact, are all chosen to match that of Fig.~\ref{fig:GP_g}. The interparticle interaction strength is kept constant during the entire cycle at the same value of the low energy equilibrium state, $c_l$, in Fig.~\ref{fig:GP_g}. This is done so that we may make a more direct comparison between the volumetric and interaction-driven quantum Otto engine cycles.

In contrast to the interaction-driven cycle, we observe an increase in the beneficial net work of the volumetric engine cycle in 3D when compared with that seen in the lower dimensions. 
This is simply explained by the dependence of the atom position variance, given in Eq.~\eqref{eq:pos_variance}, on the harmonic trapping frequency. In detail, this factor scales with the harmonic trapping frequency as $\langle \mathbf{r}^2 \rangle\propto \omega^{-4/(d+2)}$, which improves with increasing dimension. Beneficial net work extraction, which relies on $\langle \mathbf{r}^2 \rangle_h\!-\! \langle \mathbf{r}^2 \rangle_l$, is therefore also improved with increasing dimension.

Finally, we highlight that the upper bound on efficiency, given in Eq.~\eqref{eq:efficiency_bound}, for this volumetric sudden quench engine cycle is
\begin{equation}
    \eta < 1-\frac{\omega_l^2}{\omega_h^2}.
\end{equation}
Similar bounds have been derived previously for sudden quench Otto engine cycles which leverage control over one-body operators. 
In particular, it was shown in Ref.~\cite{jaramillo2016quantum} that the efficiency of the Otto heat engine is bounded from above by $\eta \!\leq\!(1-\omega_l^2/\omega_h^2)/2$, which is exactly half of our upper bound. 
However, such a bound applies generally only for non-interacting quantum gases, or gases where the interaction term in the Hamiltonian takes a scaling form \cite{husimi}, which is in general not satisfied for an ultracold Bose gas with contact interactions. We therefore again highlight that the above bound is applicable for control over \textit{any} operator with arbitrary system-reservoir contact, and is therefore universal.

\section{Conclusions}

In this work, we have generalized a previous study on the interaction-driven quantum Otto engine cycle, done in Ref.~\cite{watson2025quantum} for the uniform 1D Bose gas, to arbitrary many-body interacting quantum models. In particular, we showed in Section \ref{sec:Otto} that the net work is universally determined by the difference in the total correlation of the high and low energy equilibrium states, for Otto engine cycles where the work strokes consist of a sudden quench and the equilibration strokes allow for arbitrary system-reservoir contact. Additionally, in the limit of an infinitesimal quench of interaction strength, we demonstrated that the sudden quench engine cycle operates with the same performance as the quasi-static cycle, a fact that was previously noted for the transverse field Ising model in Ref.~\cite{sajitha2025quantum}, and here extended to arbitrary interacting quantum systems.

The efficiency of the interaction-driven cycle was seen to give rise to a universal upper bound, which takes the form $\eta \!< \!1 \!-\! c_l/c_h$ for some externally controllable interaction strength $c$. As this formula only makes reference to the ratio of interaction strengths, it is universally applicable to any sudden quench Otto engine cycle operating with such a ratio of interaction strengths.

To demonstrate the breadth of application of the formalism introduced, in Section \ref{sec:applications_interaction} we investigated operation of the interaction-driven Otto engine cycle in a variety of models. In particular, we first examined a sudden quench Otto engine cycle in the Lieb-Liniger model of a uniform 1D Bose gas, where the efficiency under an additional exchange of particles between the working fluid and the reservoirs was shown to exceed the adiabatic limit of the Otto engine cycle where \emph{only} heat was allowed to flow between the reservoirs and the working fluid. We then applied this formalism to the case of an ultracold Bose gas in $1$, $2$, and $3$ dimensions, where the breadth of engine operation with interaction strength ratio was enhanced in lower dimensions.

Furthermore, we have introduced a novel interaction-driven engine model for a 1D gas of spin-$1/2$ fermions, described by the Yang-Gaudin model. This nonequilibrium Otto engine cycle was shown to operate as purely a heat engine over a limited range of the ratio of interaction strengths, generating a small amount of beneficial net work. However, we then demonstrated how control over the spin polarization of the high and low energy thermal equilibrium states may be utilized to provide an additional source of `fuel' for its operation. Such control was shown to greatly enhance the net work, which itself depends on the inter-spin correlation. 

Finally, in Section \ref{sec:generalise_otto} we demonstrated how the methods introduced to evaluate the performance of the interaction-driven Otto engine cycle are applicable to control over arbitrary Hamiltonian operators. 
Operation of this general quantum Otto engine cycle was demonstrated for the particular case of a harmonically trapped ultra-cold Bose gas, where external control is over the harmonic trapping frequency. In contrast to the interaction-driven case, we observed that beneficial net work extraction increases with increasing dimension, a fact that was explained by the dependence of the net work on the quantity $\langle \mathbf{r}^2\rangle$, which scales favorably with the trapping frequency with increasing dimension.

As the sudden quench approximation utilized in this work is independent of the type of operator that is externally controlled, the results derived here are not limited to simple one- or two-body operators; indeed such general considerations are applicable without modification to systems with non-local interactions or higher many-body operators.
Therefore, we emphasize that all results derived for the interaction-driven cycle apply to the case of arbitrary operators, including the efficiency upper bound and quasi-static operation under infinitesimal quenches.

\section*{Acknowledgements}
 This work was supported through Australian Research Council Discovery Project Grant Nos. DP190101515 and DP240101033.\\


%

\end{document}